\documentclass[11pt,a4paper]{article}
\usepackage{amsmath,amssymb}
\usepackage{epsfig,graphicx}
\usepackage{cite}

\begin{document}

\title{\bf Dirac neutrino magnetic moment and the shock wave revival 
in a supernova explosion}

\author{A.~V.~Kuznetsov\footnote{{\bf e-mail}: avkuzn@uniyar.ac.ru},
N.~V.~Mikheev\footnote{{\bf e-mail}: mikheev@uniyar.ac.ru},
A.~A.~Okrugin\footnote{{\bf e-mail}: alexander.okrugin@gmail.com}
\\
\small{\em Yaroslavl State P.G.~Demidov University} \\
\small{\em Sovietskaya 14, 150000 Yaroslavl, Russian Federation}
}
\date{}

\maketitle

\begin{abstract}
The process of the two-step conversion of the neutrino helicity, 
$\nu_L \to \nu_R \to \nu_L$, is analysed in the supernova conditions, 
where the first stage is realized due to the interaction of the 
neutrino magnetic moment with the plasma electrons and protons 
in the supernova core. The second stage is caused by the neutrino 
resonant spin-flip in a magnetic field of the supernova envelope. 
Given the neutrino magnetic moment within the interval
$10^{-13} \, \mu_{\rm B} < \mu_\nu < 10^{-12} \, \mu_{\rm B}$, and 
with the existence of the magnetic field at the scale 
$\sim 10^{13}$ G between the neutrinosphere and the shock-wave stagnation 
region, it is shown that an additional energy of the order of 
$10^{51}$ erg can be injected into this region during the typical time 
of the shock-wave stagnation. This energy 
could be sufficient for stumulation of the damped shock wave.
%\\
%PACS: 13.15.+g, 14.60.St, 97.60.Bw
\end{abstract}

%%%%%%%%%%%%%%%%%%%%%%%%%%%%%%%%%%%%%%%%%%%%%%%%%%%%%%%%%%%%%%%%%%%%%%%%%%%%%%%%
\section{Introduction}
%%%%%%%%%%%%%%%%%%%%%%%%%%%%%%%%%%%%%%%%%%%%%%%%%%%%%%%%%%%%%%%%%%%%%%%%%%%%%%%%

\def\D{\mathrm{d}} 
\def\E{\mathrm{e}}
\def\I{\mathrm{i}}

In a modelling of the supernova explosion, two main problems 
arise~\cite{Imshennik:1988,Bethe:1990,Raffelt:1996,
Buras:2005,Janka:2007}. First, the mechanism of the damped shock wave
stimulation has not been developed completely yet. It is believed that the 
explosion cannot be realized without the shock wave revival.	
Let us remind, that the main reason of the shock-wave damping is the energy 
loss by the nuclei dissociation. 
The second problem is that even in the case of the ``successful''
theoretical supernova explosion, the energy release turns out to be 
essentialy less than the observed kinetic energy of the envelope
$\sim 10^{51}$ erg. That is known as the FOE problem 
(ten to the Fifty One Ergs). 
Thus, it is necessary for the self-consistent description 
of the explosion dynamics, that the neutrino flux, outgoing from
the supernova core, could transfer by some mechanism the energy 
$\sim 10^{51}$ erg to the supernova envelope.

A possible solution of those problems, 
first proposed by A.~Dar~\cite{Dar:1987}, was based on the
assumption of the existence of the neutrino magnetic moment being not too small. 
A huge number of left-handed electron neutrinos $\nu_e$
is produced in the collapsing supernova 
core, and a part of them could convert into right-handed neutrinos 
due to the interaction of the neutrino magnetic moment with plasma electrons and protons.
These right-handed neutrinos, being sterile with respect to the weak interaction,
freely escape from the central part of the supernova, if the neutrino magnetic moment 
is not too large, $\mu_\nu < 10^{-11} \, \mu_{\rm B}$,  
where $\mu_{\rm B}$ is the Bohr magneton.
In the supernova envelope, a part of these neutrinos can flip back to the 
left-handed ones due to the interaction of the neutrino magnetic moment with 
a magnetic field. It is now believed that the magnetic field strength 
in the supernova envelope could achieve the critical value 
$B_e = m_e^2/e \simeq 4.41 \times 10^{13}$ G~\footnote{We use
the natural system of units $c = \hbar = 1$. $e > 0$ is an elementary charge.} 
and even exceed it. 
The produced left-handed neutrinos, being absorbed in beta-processes, 
$\nu_e n \to e^- p$,
can transfer an additional energy to the supernova envelope. 

In our opinion, a reason arises at the present time to reconsider in more 
detail the Dar's mechanism. In the recent paper~\cite{Kuznetsov:2007} 
we have shown that the evaluations of the right-handed neutrino flux and the 
luminosity from the supernova core were essentially understated in the 
previous papers on the subject. 

In this paper, we perform an analysis of the two-step conversion of the 
neutrino helicity,
$\nu_L \to \nu_R \to \nu_L$, under the supernova conditions, 
and of the possibility of the damped shock wave stumulation by this process.

%%%%%%%%%%%%%%%%%%%%%%%%%%%%%%%%%%%%%%%%%%%%%%%%%%%%%%%%%%%%%%%%%%%%%%
\section{The right-handed neutrino luminosity} 
\label{sec:nu_R_luminosity}
%%%%%%%%%%%%%%%%%%%%%%%%%%%%%%%%%%%%%%%%%%%%%%%%%%%%%%%%%%%%%%%%%%%%%%

The process of the neutrino chirality flip $\nu_L \to \nu_R$
under the physical conditions of the supernova core was investigated in the 
papers~\cite{Barbieri:1988,Ayala:1999,Ayala:2000,Kuznetsov:2007}.
The process is possible due to the interaction of the Dirac neutrino magnetic 
moment with a virtual plasmon, which can be both produced and absorbed: 
$$\nu_L \to \nu_R + \gamma^*; \quad \nu_L + \gamma^* \to \nu_R.$$
The detailed calculation of the plasma polarization effect 
on the photon propagator reveals, in particular, 
that the contribution of the proton component of plasma is dominant. 
As a result a new astrophysical bound on the electron-type neutrino magnetic 
moment was established~\cite{Kuznetsov:2007} from the supernova $SN1987A$ data:

\begin{eqnarray}
\mu_\nu < (0.7 - 1.5) \, \times 10^{-12} \, \mu_{\rm B}\,,
\label{eq:mu_fr_Q}
\end{eqnarray}

which improved the existed constraint by the factor of 2.

Particularly, a function $\Gamma_{\nu_R} (E)$ defining the spectrum
of the right-handed neutrino energies, was calculated 
in Ref.~\cite{Kuznetsov:2007}.
In other words, this function defines the number of right-handed neutrinos
emitted per 1 MeV of the neutrino energy spectrum, per unit time, 
from the unit volume of the supernova core:

\begin{eqnarray}
\frac{\mathrm{d} n_{\nu_R}}{\mathrm{d} E} = 
\frac{E^2}{2 \, \pi^2} \, \Gamma_{\nu_R} (E) \,.
\label{eq:dn/dE}
\end{eqnarray}

The function $\Gamma_{\nu_R} (E)$ also determines the spectral density
of the right-handed neutrino luminosity of the supernova core:

\begin{eqnarray}
\frac{\mathrm{d} L_{\nu_R}}{\mathrm{d} E}
 = V\, \frac{\mathrm{d} n_{\nu_R}}{\mathrm{d} E} \, E = 
V \, \frac{E^3}{2 \, \pi^2} \, \Gamma_{\nu_R} (E) \,,
\label{eq:dL/dE}
\end{eqnarray}

where $V$ is the volume of the area emitting neutrinos. 

\begin{figure}
\begin{center}
\includegraphics*[width=0.9\textwidth]{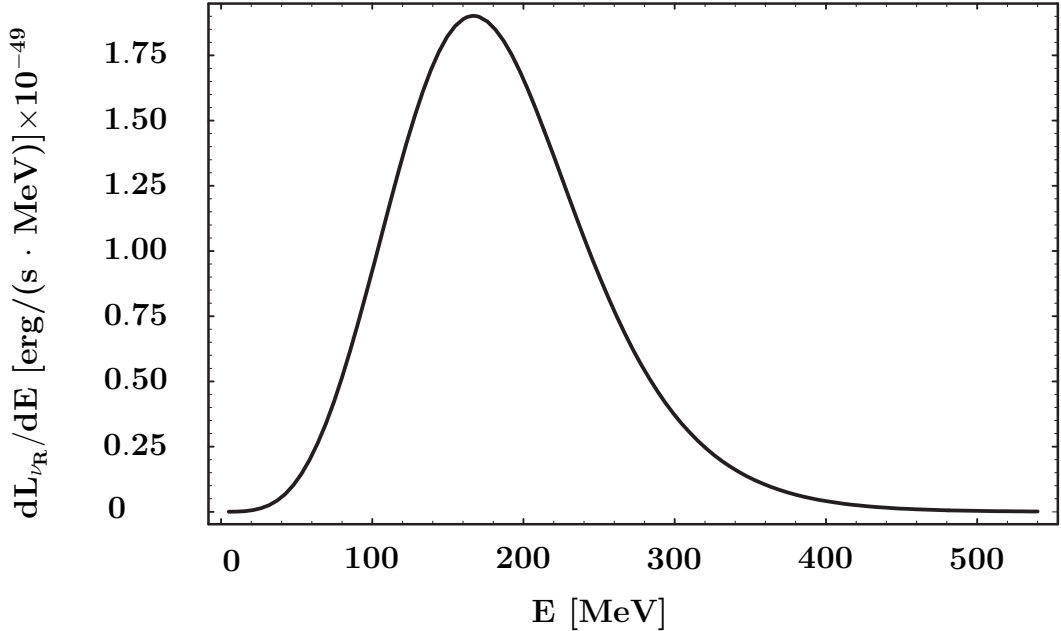}
\end{center}
\caption{The energy spectrum of the right-handed neutrino 
luminosity for the plasma temperature $T =$ 30 MeV and  
for $\mu_\nu = 3 \times 10^{-13} \, \mu_{\rm B}$.} 
\label{fig:emissivity}
\end{figure}

The function $\mathrm{d} L_{\nu_R}/\mathrm{d} E$ was calculated 
in Ref.~\cite{Kuznetsov:2007}, and it is shown 
in the figure~\ref{fig:emissivity} 
for the typical supernova core parameter values: 
the temperature $T \simeq$ 30 MeV, the electron and neutrino chemical potentials
$\tilde \mu_e \simeq$ 300 MeV, $\tilde \mu_{\nu_e} \simeq$ 160 MeV,
the volume $V \simeq 4 \times 10^{18} \, \mbox{cm}^3$
and for the neutrino magnetic moment $\mu_\nu = 3 \times 10^{-13} \, \mu_{\rm B}$.

%%%%%%%%%%%%%%%%%%%%%%%%%%%%%%%%%%%%%%%%%%%%%%%%%%%%%%%%%%%%%%%%%%%%%%
%%%%%%%%%%%%%%%%%%%%%%%%%%%%%%%%%%%%%%%%%%%%%%%%%%%%%%%%%%%%%%%%%%%%%%

The integral luminosity of the right-handed neutrinos appeared to be the 
following:

\begin{eqnarray}
L_{\nu_R} = 4 \times 10^{51} \, \frac{\mbox{erg}}{\mbox{s}} \,.
\label{eq:L}
\end{eqnarray}

Hereafter we use for the definiteness the neutrino magnetic moment value 
$\mu_\nu = 3 \times 10^{-13} \, \mu_{\rm B}$.
On the one hand, this value is sufficiently small to avoid a distortion of 
the supernova dynamics. On the other hand, it is large enough to provide 
the required level of the luminosity~(\ref{eq:L}).

If the right-handed neutrino energy was converted 
into the energy of the left-handed neutrinos, for example due to the
well-known mechanism of the spin oscillations, then during the typical stagnation time
of the shock wave of the order of some tenths of a second, 
an additional energy of order $10^{51}$ erg could be injected 
into the supernova envelope.

%%%%%%%%%%%%%%%%%%%%%%%%%%%%%%%%%%%%%%%%%%%%%%%%%%%%%%%%%%%%%%%%%%%%%%
\section{The resonant transition $\nu_R \to \nu_L$
in the magnetic field
 \protect\\ of the supernova envelope}
\label{sec:resonance}
%%%%%%%%%%%%%%%%%%%%%%%%%%%%%%%%%%%%%%%%%%%%%%%%%%%%%%%%%%%%%%%%%%%%%%

We consider a part of the supernova envelope between the neutrinosphere
(of the radius $R_\nu$) and the shock wave stagnation region 
(of the radius $R_s$). By the present conceptions, 
typical values of $R_\nu$ and $R_s$ vary rather slightly during the stagnation time. 
These values could be estimated as $R_\nu \sim$ 20--50~km, $R_s \sim$ 100--200~km.
If a sufficiently large magnetic field $\sim 10^{13}$ G exists in this region, 
then the spin oscillation phenomenon takes place, 
which can be of the resonant type at certain conditions. 

It is convenient to illustrate the magnetic field influence on a neutrino 
with a magnetic moment by means of the equation of the neutrino helicity 
evolution in an external uniform magnetic field. 
Taking into account the additional energy $C_L$, which 
the left-handed electron type neutrino $\nu_e$ 
acquires in medium, the equation of the helicity evolution can be written in the 
form~\cite{Voloshin:1986a,Voloshin:1986b,Okun:1986,Voloshin:1986c,Okun:1988} 

\begin{equation}
{\mathrm i}\,\frac{\partial}{\partial t}
\left( 
\begin{array}{c} 
\nu_R \\ \nu_L 
\end{array} 
\right)
=
\left[\hat E_0 +
\left( 
\begin{array}{cc} 
0 & \mu_\nu B_{\perp} \\ \mu_\nu B_{\perp} & C_L
\end{array} 
\right)
\right]
\left( 
\begin{array}{c} 
\nu_R \\ \nu_L 
\end{array} 
\right) \,,
\label{eq:evolution} 
\end{equation}

where

\begin{equation}
C_L = \frac{3 \, G_{\mathrm F}}{\sqrt{2}} \, \frac{\rho}{m_N} 
\left( Y_e + \frac{4}{3} \, Y_{\nu_e} - \frac{1}{3} \right) \,.
\label{eq:C_L}
\end{equation}

Here, the ratio $\rho/m_N = n_B$ is the nucleon density,
$Y_e = n_e/n_B = n_p/n_B, \, Y_{\nu_e} = n_{\nu_e}/n_B$, 
$n_{e,p,\nu_e}$ are the densities of electrons, protons and neutrino respectively.
$B_{\perp}$ is the transverse component of the magnetic field with respect to 
the neutrino movement direction, $\hat E_0$ is proportional to the unit 
matrix and is inessential for our analysis. 

\begin{figure}
\begin{center}
\includegraphics*[width=0.7\textwidth]{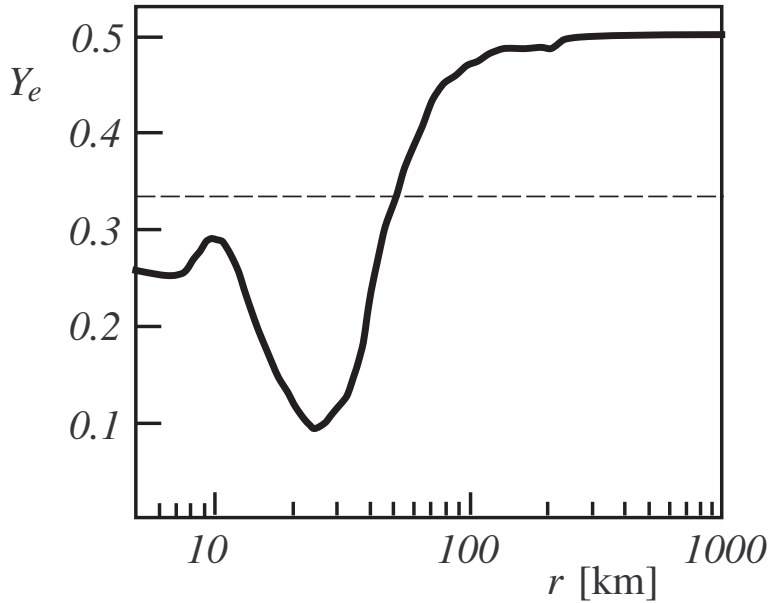}
\end{center}
\caption{The qualitative behaviour of the dependence 
$Y_e (r)$ about 0.1 to 0.2 s after the shock formation, 
with the typical gap caused by the ``short'' neutrino outburst, 
see e.g. {\it Buras et al., 2005}. The dashed line corresponds to the value 
$Y_e = 1/3$.} 
\label{fig:gap}
\end{figure}

The expression~(\ref{eq:C_L}) for the additional energy of left-handed 
neutrinos $C_L$ deserves a special analysis. 
It is remarkable that the possibility exists for this value to be zero 
just in the region of the supernova envelope we are interested in. 
And in turn this is the condition of the resonant transition $\nu_R \to \nu_L$.
Taking into account that the neutrino density in the supernova envelope 
is sufficiently small, one may neglect the value $Y_{\nu_e}$ 
in the expression~(\ref{eq:C_L}), that gives the condition of the resonance 
in the form $Y_e = 1/3$. It should be noted that the values $Y_e$ which 
are realized in the supernova envelope, typical for the collapsing 
matter, are: $Y_e \sim$ 0.4--0.5. 
However, the shock wave causes the nuclei dissociation and 
makes the substance to be more transparent for neutrinos. 
This leads to the so-called ``short'' neutrino outburst
and consequently to the significiant matter deleptonization in this region.
According to the existing conceptions, a typical gap arises in the radial 
distribution of the value $Y_e$, where $Y_e$ may fall down to 
the value $\sim 0.1$, see, for example~\cite{Bethe:1990,Buras:2005}.
The qualitative behaviour of the dependence 
$Y_e (r)$ is represented in the figure~\ref{fig:gap}. 
Thus, a point necessarily exists where $Y_e$ takes the value of $1/3$.
It is remarkable, that only one such point appears, 
with $\mathrm{d} Y_e / \mathrm{d} r > 0$,
see~\cite{Bethe:1990,Buras:2005}. 

%%%%%%%%%%%%%%%%%%%%%%%%%%%%%%%%%%%%%%%%%%%%%%%%%%%%%%%%%%%%%%%%%%%%%%
%%%%%%%%%%%%%%%%%%%%%%%%%%%%%%%%%%%%%%%%%%%%%%%%%%%%%%%%%%%%%%%%%%%%%%

Notice, that the condition $Y_e = 1/3$ is the necessary but still not 
the sufficient one for the resonant conversion $\nu_R \to \nu_L$.
The realization is also necessary of the so-called adiabatic condition. 
This means that the diagonal element $C_L$ in the equation~(\ref{eq:evolution}),
at least, should not exceed the nondiagonal element $\mu_\nu B_{\perp}$, 
when the shift is made from the resonance point at the distance 
of the order of the oscillations length. This leads to the 
condition~\cite{Voloshin:1988}: 

\begin{eqnarray}
\mu_\nu B_{\perp} \gtrsim \left( \frac{\mathrm{d} C_L}{\mathrm{d} r} \right)^{1/2} 
\simeq \left( \frac{3 \, G_{\mathrm F}}{\sqrt{2}} \, \frac{\rho}{m_N} \, 
\frac{\mathrm{d} Y_e}{\mathrm{d} r}\right)^{1/2} .
\label{eq:res_cond}
\end{eqnarray}

The typical parameter values in the considered area 
are the following, see~\cite{Bethe:1990,Buras:2005}:

\begin{eqnarray}
\frac{\mathrm{d} Y_e}{\mathrm{d} r} \sim 10^{-8} \, \mbox{cm}^{-1} \,, 
\quad
\rho \sim 10^{10} \, \mbox{g} \cdot \mbox{cm}^{-3} \,.
\label{eq:param}
\end{eqnarray}

For the magnetic field value, providing the realization of the resonance 
condition, one can find:

\begin{eqnarray}
B_{\perp} \gtrsim 2.6 \times 10^{13} \mbox{G} 
\left( 
\frac{10^{-13} \mu_{\rm B}}{\mu_\nu} \right) 
\left( \frac{\rho}{10^{10} \mbox{g} \cdot \mbox{cm}^{-3}}
\right)^{1/2}
\left( \frac{\mathrm{d} Y_e}{\mathrm{d} r} \times 10^8 \, \mbox{cm}
\right)^{1/2} .
\label{eq:res_cond_B}
\end{eqnarray}

Thus, the performed analysis shows that the Dar scenario 
of the two-step conversion of the neutrino helicity,
$\nu_L \to \nu_R \to \nu_L$, can be realized, 
if the value of the neutrino magnetic moment is in the interval

\begin{eqnarray}
10^{-13} \, \mu_{\rm B} < \mu_\nu < 10^{-12} \, \mu_{\rm B} \,,
\label{eq:munu_int}
\end{eqnarray}

and under the condition that the magnetic field of the scale 
$10^{13}$ G exists in the region $R_\nu < R < R_s$. 
During the shock wave stagnation time $\Delta t \sim$ 0.2--0.4 sec
the additional energy can be injected into this region, of the order of

\begin{eqnarray}
\Delta E \simeq L_{\nu_R} \, \Delta t \sim 10^{51} \, \mbox{erg} \,,
\label{eq:DeltaE}
\end{eqnarray}

which is just enough for the problem solution. 

%%%%%%%%%%%%%%%%%%%%%%%%%%%%%%%%%%%%%%%%%%%%%%%%%%%%%%%%%%%%%%%%%%%%%%%%%%%%%%%%
\section{Conclusion}
%%%%%%%%%%%%%%%%%%%%%%%%%%%%%%%%%%%%%%%%%%%%%%%%%%%%%%%%%%%%%%%%%%%%%%%%%%%%%%%%

We have re-analysed the two-step conversion of the 
neutrino helicity, $\nu_L \to \nu_R \to \nu_L$, under the supernova 
conditions. As we have shown, this conversion process could provide 
an additional energy of the order of 
$10^{51}$ erg which can be injected into the region between the 
neutrinosphere and the shock-wave stagnation area, $R_\nu < R < R_s$, 
during the typical stagnation time of the order of some tenths of a second. 
This energy could be sufficient for stumulation of the damped shock wave. 

The conditions for the realization of this scenario appear to be not very 
rigid. 
The Dirac neutrino magnetic moment should belong to the interval
$10^{-13} \, \mu_{\rm B} < \mu_\nu < 10^{-12} \, \mu_{\rm B}$, and 
the magnetic field $\sim 10^{13}$ G should exist in the region 
$R_\nu < R < R_s$. 

%%%%%%%%%%%%%%%%%%%%%%%%%%%%%%%%%%%%%%%%%%%%%%%%%%%%%%%%%%%%%%%%%%%%%%%%%%%%%%%%
\section*{Acknowledgments}
%%%%%%%%%%%%%%%%%%%%%%%%%%%%%%%%%%%%%%%%%%%%%%%%%%%%%%%%%%%%%%%%%%%%%%%%%%%%%%%%

The authors are grateful to M.~I.~Vysotsky for useful discussion. 
A.~K. and N.~M. express their deep gratitude to the organizers of the 
Seminar ``Quarks-2008'' for warm hospitality.

The work was supported in part by the Russian Foundation for Basic Research 
under the Grant No.~07-02-00285-a, and by the Council on Grants by the President
of the Russian Federation for the Support of Young Russian Scientists
and Leading Scientific Schools of Russian Federation under the Grant 
No.~NSh-497.2008.2.

%%%%%%%%%%%%%%%%%%%%%%%%%%%%%%%%%%%%%%%%%%%%%%%%%%%%%%%%%%%%%%%%%%%%%%%%%

%%%%%%%%%%%%%%%%%%%%%%%%%%%%%%%%%%%%%%%%%%%%%%%%%%%%%%%%%%%%%%%%%%%%%%%%%

\end{document}